\documentclass[12pt,letterpaper,copyedit]{article}
\usepackage{osajnl2}
\usepackage[draft]{hyperref}
\usepackage{amsmath}
\usepackage{soul}
\usepackage{graphics,graphicx,epsfig}

\begin{document}
\title{Theory of Polarization Attraction in Parametric Amplifiers Based on Telecommunication Fibers}
\author{Massimiliano Guasoni$^1$, Victor V. Kozlov$^{1,2}$, and Stefan Wabnitz$^{1}$}
\affiliation{$^1$Department of Information Engineering, Universit\`{a} di
Brescia,\\ Via Branze 38, 25123 Brescia, Italy}
\affiliation{$^2$Department of Physics, St.-Petersburg State University,\\
Petrodvoretz, St.-Petersburg, 198504, Russia}

\affiliation{$^*$Corresponding author: massimiliano.guasoni@ing.unibs.it}

\begin{abstract}
We develop from first principles the coupled wave equations that describe polarization-sensitive parametric amplification based on four-wave mixing in 
standard (randomly birefringent) optical fibers. We show that in the small-signal case these equations can be solved analytically, and permit us to predict
the gain experienced by the signal beam as well as its state of polarization (SOP) at the fiber output. We find that, independently of its initial value, the
output SOP of a signal within the parametric gain bandwidth is solely determined by the pump SOP. We call this effect of pulling the polarization of the
signal towards a reference SOP as polarization attraction, and such parametric amplifier as the FWM-polarizer. Our theory is valid beyond the zero 
polarization mode dispersion (PMD) limit, and it takes into account moderate deviations of the PMD from zero. In particular, our theory is capable of
analytically predicting the rate of degradation of the efficiency of the parametric amplifier which is caused by the detrimental PMD effect.   
\end{abstract}


\maketitle
%===============
\section{Introduction}
Recent years have witnessed a substantial growth of interest in developing nonlinear-optical techniques for the control of the state of polarization (SOP)
of light beams. The motivation behind such research activities is twofold. First of all, nonlinear optical techniques may permit replacing the inefficient and
lossy method of polarizing a light beam by conventional passive linear polarizers with a the lossless polarization attraction of an arbitrary initial SOP
towards the desired SOP at the output of a nonlinear medium. A key advantage of using lossless polarization attraction is that, in contrast with passive
linear polarizers, input signal SOP changes do not lead to output signal intensity fluctuations or relative intensity noise (RIN). The second goal is to find
efficient ways to excercise all-optical control over the SOP of a signal beam by exploiting its nonlinear interaction with a pump beam with a 
well-determined SOP. Here we analyse a novel method for achieving the all-optical control of the SOP of a signal beam, namely exploiting the 
four-wave-mixing-mediated process of parametric amplification in a standard telecom optical fiber.

In short, nonlinear-optical methods allow for designing novel types of polarizers with much greater functionality than conventional passive linear 
polarizers. So far, two distinctly different types of nonlinear-optical polarizers were proposed. The first class comprises the so-called 
nonlinear lossless polarizers (NLPs), which are based on the cross-polarization modulation (XPolM) of two intense beams in a Kerr medium. To the 
second class belong the so-called Raman polarizers, which are based on the polarization-sensitive Raman amplification of a signal beam in a 
Raman-active medium. These two types of polarizers exploit the two complementary manifestations of the cubic nonlinearity of fibers -- conservative for 
inducing XPolM effect, and dissipative which is responsible for the Raman effect. Here, we exploit the same cubic nonlinearity, more precisely its 
conservative part, for initiating the process of polarization-sensitive four-wave mixing (FWM) between three beams. 

The first NLP was proposed and experimentally demonstrated by Heebner {\it et al} in Ref.~\cite{boyd}. It was based not on the Kerr nonlinearity, but on 
a photorefractive effect. This polarizer was capable of transforming, in a lossless manner, a light beam with an arbitrary initial SOP into a beam with one 
and the same SOP towards its output. The principle of operation of this device was the conversion of energy from one polarization component of the
beam into its orthogonal polarization component. Photorefractive materials are characterized by a nonlinear response which is far too slow to be useful in
contemporary ultrafast optics. In contrast, the Kerr nonlinearity of silica is virtually instantaneous, which makes optical fibers a promising medium for 
implementing lossless polarizers within high-bit-rate telecom networks. The progress in developing fiber-based NLPs started from impractical isotropic 
fibers \cite{Pitois1,Pitois2,EPL} and evolved towards cheap and reliable telecom fibers \cite{FatomeOE,Kozlov2011,Kozlov_coprop} or specialty fibers
such as highly-birefringent and spun fibers \cite{KozlovOL}. The mathematical aspects of the problem were studied in
Refs.~\cite{torus1,torus2,torus3,torus4,torus5}, and allowed to get further insight into the physics of fiber-based NLPs, whose principle of operation is 
different from that of photorefractive lossless polarizers. Instead of the self-interaction of a single beam in a photorefractive material, a two-beam 
cross-interaction (namely, XPolM) is used in the Kerr medium. Namely, an auxiliary pump beam with a well-defined SOP is employed, serving as a 
polarization reference for the signal beam with arbitrary initial SOP. As previously outlined, when using lossless polarizers input signal SOP fluctuations 
do not lead to output RIN \cite{FatomeOE}.

Another type of nonlinear-optical polarizer is the Raman polarizer. It is different from conventional Raman amplifiers by its sensitivity to the SOP of the 
pump beam. The signal which experiences Raman amplification acquires an SOP which is dictated by the SOP of the pump. In this way we may 
exercise an all-optical control over the polarization of the signal beam. Note that conventional fiber-optic Raman amplifiers operate in the regime where 
the output SOP of the signal is independent on the pump SOP. The first Raman polarizer was demonstrated by Martinelli {\it et. al.} in 
Ref.~\cite{martinelli}, followed by a number of theoretical papers \cite{our1,KozlovJLT,padova1,padova2,ourPTL,Sergeyev,Springer}. Similar
polarization-sensitive amplification was predicted theoretically and confirmed experimentally in Ref.~\cite{Zadok} for the Brillouin amplification of a signal
beam in standard optical fibers. These devices can be similarly called Brillouin polarizers. Since they are based on a gain mechanism, which is 
maximum whenever the signal and pump SOPs are aligned and zero when they are orthogonal, in general both Raman and Brillouin-based polarizers 
suffer from severe output RIN in the presence of input signal SOP fluctuations. 

A common feature uniting all of these nonlinear fiber-optic polarizers is that they can operate efficiently only in the limit of vanishing polarization mode 
dispersion (PMD). PMD is the effect which is caused by random variations of the magnitude and/or orientation of the birefringence along the fiber length, 
and it is acquired as a result of inevitable technical imperfectnesses in the process of drawing a fiber from a preform. Recent progress in fiber 
manufacturing brought to the market fibers with much lower values of PMD than it was previously available. It is this technological breakthrough that 
made it possible the observation of the previously discussed polarization-sensitive effects in optical fibers. Theoretical estimates show that the smaller the 
PMD coefficient, the shorter the total fiber length and the smaller the frequency separation of the signal and the pump beams, the better the performance 
of all of the above described polarizers. It is one of the main goals of a theory to be able to predict the degradation rate of useful polarization attraction 
effects which is caused by PMD. Such degradation rates for NLPs and Raman polarizers have been calculated analytically in
Refs.~\cite{Agrawal_Raman,Springer}. 

It is important to note that the concept of all of these smart polarizers is not limited to fiber-optics applications only. Indeed, nonlinear polarizers can be 
implemented with any optical waveguide exhibiting Kerr and/or Raman nonlinearity. Using integrated optics waveguides may lift the problems which are 
associated with fiber PMD, and even make nonlinear polarizers less bulky and more compact, providing that the waveguide material exhibits nonlinear 
coefficients which are much larger than silica. For example, the silicon-based Raman polarizer proposed in Ref.~\cite{silicon} is free of the PMD-induced 
degradation and has a cm-long size as compared to the km-long fibers, thanks to three-four orders of magnitude Raman gain enhancement in silicon 
with respect to silica.

The present theoretical study extends the concept of nonlinear polarizers to the FWM process in telecom fibers. The goal here is to find the conditions 
upon which the process of parametric amplification is sensitive to the SOP of the pump beam. In this way we arrive to the notion of a FWM-polarizer, 
meaning that the SOP of the amplified signal beam is determined by the SOP of the fully polarized pump beam. We derive here the coupled wave 
equations for the pump, idler, and signal beams. In the limit of zero PMD, these equations reduce to the equations which were previously derived by 
McKinstrie {\it et. al.} in Ref.~\cite{mckinstrie2004} for describing degenerate FWM in standard fibers. The major advantage of our theory is its applicability 
(slightly) beyond the zero-PMD limit, in the sense that it is capable of predicting the degradation rate of the efficiency of the FWM-polarizer for low-PMD 
fibers as well. Knowing this degradation rate allows one to properly design practical fiber based nonlinear polarizers. The present work substantially 
extends to the case of random birefringence telecom fibers a previous study of polarization attraction in deterministic, high-birefringence optical fibers 
~\cite{guaso2011}. Note that the polarization-sensitive parametric amplification in optical fibers was studied theoretically by Lin and Agrawal 
\cite{agrawal2004,agrawalOL}, and also theoretically and experimentally by Freitas {\it et. al.} in Ref.~\cite{Freitas2007} and resulted in a proposal of a
fiber-based polarization switch. As discussed in Ref.~\cite{guaso2011}, FWM-based polarizers are based on the polarization sensitivity of parametric
gain. Such gain is maximum for a signal SOP which is aligned with that of the pump, and zero for a signal SOP orthogonal to the pump. Thus 
FWM-polarizers are not immune from output RIN resulting from input signal polarization fluctuations. Nevertheless, since parametric gain is generally 
larger than Raman gain in silica fibers, FWM-polarizers may employ shorter fibers or lower pump powers than Raman polarizers. In addition, since the 
repolarization capability of FWM polarizers is based on parametric gain, these devices provide a more flexible control over the gain and repolarization 
bandwidth. Indeed, such bandwidth may be extended up to 70 nm and even include the normal dispersion regime by properly engineering the 
wavelength dependence of the fiber dispersion and by adjusting the pump power~\cite{WabnitzFOPA}.

%========================
\section{Equations of the model}
We shall consider the process of degenerate four-wave-mixing (FWM). This process involves three waves with frequencies that satisfy the matching 
condition $2\omega_p =\omega_s+\omega_i$. Pump, signal, and idler waves are labeled correspondingly as $p,\, s,\, i$. All three waves are 
co-propagating along the $z$ direction in a telecommunication (i.e. randomly birefringent) fiber. The vectorial theory of parametric amplification in fibers
was developed in Refs.~\cite{agrawal2004,agrawalOL,mckinstrie2004}, basing on the tensorial properties of silica in the telecom band. The starting 
equation is derived under standard for nonlinear optics approximations, from Maxwell's equation with a polarization which takes into account the 
nonlinear cubic response of silica and the birefringence of the fiber. Utilizing the Jones representation and transforming into a reference frame moving 
with the speed of light at the corresponding frequency $\omega_f$ (with $f=\{ p,s,i\})$), the equations for the Jones vectors of the pump and the signal 
read as
\begin{eqnarray}
& & i\frac{\partial U_p}{\partial z} + \Delta B(\omega_p,\, z) U_p
\nonumber\\
& & + \frac{2}{3}\gamma[(U_p\cdot U_p^*)U_p + \frac{1}{2}(U_p\cdot U_p)U_p^*]
\nonumber\\
& & + \frac{2}{3}\gamma[(U_s\cdot U_s^*)U_p+(U_s\cdot U_p)U_s^*+(U_p\cdot U_s^*)U_s+(U_i\cdot U_i^*)U_p+(U_i\cdot U_p)U_i^*
+(U_p\cdot U_i^*)U_i]
\nonumber\\
& & + \frac{2}{3}\gamma \exp(i\Delta k z)[(U_i\cdot U_s)U_p^*+(U_i\cdot U_p^*)U_s+(U_s\cdot U_p^*)U_i ]=0\, ,
\label{1}
\end{eqnarray}
\begin{eqnarray}
& & i\frac{\partial U_s}{\partial z} + \Delta B(\omega_s,\, z) U_s + iB'_s \frac{\partial U_s}{\partial t}
\nonumber\\
& & +\frac{2}{3}\gamma[(U_s\cdot U_s^*)U_s + \frac{1}{2}(U_s\cdot U_s)U_s^*] 
\nonumber\\
& & +\frac{2}{3}\gamma[(U_p\cdot U_p^*)U_s+(U_p\cdot U_s)U_p^*+(U_s\cdot U_p^*)U_p+(U_i\cdot U_i^*)U_s+(U_i\cdot U_s)U_i^*
+(U_s\cdot U_i^*)U_i]
\nonumber\\
& & + \frac{2}{3}\gamma \exp(-i\Delta k z)[ \frac{1}{2}(U_p\cdot U_p)U_i^*+(U_p\cdot U_i^*)U_p ]=0\, ,
\label{2}
\end{eqnarray}
while the idler equation is obtained from Eq.~(\ref{2}) by exchanging labels $s$ and $i$.

The Jones vectors $U_f=(u_{xf},u_{yf})^T$ are two-component vectors with $u_{xf}(z,t)$ and $u_{yf}(z,t)$ being the amplitudes of the polarization 
components in a fixed laboratory reference frame $(x,\, y)$. Note that the last lines in Eqs.~(\ref{1}) and (\ref{2}) represent the so-called energy-exchange 
terms. They are responsible for the transfer of energy between different waves, and as such are most important for our analysis of parametric 
amplification. The wave vector mismatch $\Delta k=\beta_{ei}+\beta_{es}-2\beta_{ep}=\beta_{oi}+\beta_{os}-2\beta_{op}$ should be as small as
possible for efficient energy conversion to take place. Here $\beta_{ef}$ and $\beta_{of}$ are the propagation constants of the modes aligned with 
extraordinary ($e$) and ordinary ($o$) axes at frequency $\omega_f$, respectively. The $2\times 2$ matrix 
$\Delta B(\omega_f)=\Delta\beta (\omega_f)(\cos\theta \sigma_3+\sin\theta\sigma_1)$ represents the birefringence tensor, where 
$\Delta\beta (\omega_f)$ is the value of the fiber birefringence at frequency $\omega_f$. Moreover, $\theta$ is the angle of orientation of the axis of the 
birefringence with respect to the fixed reference frame which is defined by the polarization modes ${\mathbf e}_x$ and ${\mathbf e}_y$. $\sigma_3$ and 
$\sigma_1$ are the usual Pauli matrices.

The orientation angle $\theta$ is randomly varying in fibers used for telecommunication applications, which explains the term randomly birefringent fibers 
that is applied to them. In principle, the magnitude of the birefringence $\Delta\beta (\omega_f)$ also varies stochastically along $z$. However, as noticed 
in Ref.~\cite{Wai_Menyuk}, the two approaches, one in which $\theta$ is the only stochastic variable, and the second, where both $\theta$ and 
$\Delta\beta$ are stochastic variables, produce nearly identical results. Thus, here we shall develop our theory by assuming the single stochastic
variable $\theta$. The angle $\theta$ is driven by a white noise process $\partial_z\theta =g_\theta (z)$, where $\langle g_\theta (z)\rangle =0$ and
$\langle g_\theta (z)g_\theta (z^\prime )\rangle =2L_c^{-1}\delta (z-z^\prime )$. Here $L_c$ is the correlation length which characterizes the typical 
distance over which $\theta$ changes randomly. The theory developed below is the natural extension of the one-beam theory of Wai and Menyuk in 
Ref.~\cite{Wai_Menyuk}, and two-beam theory of Kozlov, Nu$\bar{\hbox{n}}$o and Wabnitz in Ref.~\cite{Kozlov2011} to the case of three interacting
beams. All details of the derivations of the final equations of motion with deterministic coefficients starting from Eqs.~(\ref{1}), (\ref{2}) with stochastic 
coefficients, as well as the approximations which appeared on the way, can be found in Appendices A and B. Here we write down the final result:
\begin{eqnarray}
& & i\frac{\partial \phi_{1p}}{\partial z} + i\beta'_{p}\frac{\partial \phi_{1p}}{\partial t}
\nonumber\\
& & +\frac{2}{3} \left( 2 + \frac{2}{3} C_a(z) \right) \phi_{1i} \phi_{1p} \phi_{1i}^* + \frac{8}{9}  \phi_{1p}^2 \phi_{1p}^* 
\nonumber\\
& & + \frac{2}{3} \left( 2 C_a(z)  + \frac{2}{3} C_b(z) \right) \phi_{1i} \phi_{1s} \phi_{1p}^* e^{i \Delta k z} + 
 \frac{2}{3} \left( 2 + \frac{2}{3} C_a(z) \right) \phi_{1p} \phi_{1s} \phi_{1s}^* 
\nonumber\\
& & + \frac{2}{3} \left( 2 - \frac{2}{3} C_a(z) \right) \phi_{1p} \phi_{2i} \phi_{2i}^* + 
 \frac{8}{9} C_a(z)  \phi_{1i} \phi_{2p} \phi_{2i}^* 
\nonumber\\
& & + \frac{2}{3} \left( C_a(z)  + \frac{1}{3} C_b(z) \right) \phi_{1s} \phi_{2i} \phi_{2p}^*e^{i \Delta k z} + 
 \frac{8}{9} \phi_{1p} \phi_{2p} \phi_{2p}^* 
\nonumber\\
& & + \frac{2}{3} \left( C_a(z)  + \frac{1}{3} C_b(z) \right) \phi_{1i} \phi_{2s} \phi_{2p}^*e^{i \Delta k z} + 
 \frac{8}{9} C_a(z)  \phi_{1s} \phi_{2p} \phi_{2s}^* 
\nonumber\\
& & + \frac{2}{3} \left( 2 - \frac{2}{3} C_a(z) \right) \phi_{1p} \phi_{2s} \phi_{2s}^*=0
\label{3}
\end{eqnarray}
\begin{eqnarray}
& & i\frac{\partial \phi_{1s}}{\partial z} + i\beta'_{s}\frac{\partial \phi_{1s}}{\partial t}
\nonumber\\
& & +\overbrace{\frac{2}{3} \left( C_a(z)  + \frac{1}{3} C_b(z) \right) \phi_{1p}^2 \phi_{1i}^*e^{-i \Delta k z}}^{X_1} + 
 \frac{2}{3} \left( 2 + \frac{2}{3} C_a(z) \right) \phi_{1i} \phi_{1s} \phi_{1i}^* 
\nonumber\\
& & + \frac{2}{3} \left( 2 + \frac{2}{3} C_a(z) \right) \phi_{1p} \phi_{1s} \phi_{1p}^* + \frac{8}{9} \phi_{1s}^2 \phi_{1s}^* + 
 \frac{2}{3} \left( 2 - \frac{2}{3} C_a(z) \right) \phi_{1s} \phi_{2i} \phi_{2i}^* 
\nonumber\\
& & + \overbrace{\frac{2}{3} \left( C_a(z)  + \frac{1}{3} C_b(z) \right) \phi_{1p} \phi_{2p} \phi_{2i}^*e^{-i \Delta k z}}^{X_2} + 
 \frac{8}{9} C_a(z)  \phi_{1i} \phi_{2s} \phi_{2i}^* 
\nonumber\\
& & + \frac{2}{3} \left( 2 - \frac{2}{3} C_a(z) \right) \phi_{1s} \phi_{2p} \phi_{2p}^* + 
 \frac{8}{9} C_a(z)  \phi_{1p} \phi_{2s} \phi_{2p}^* + 
 \frac{8}{9} \phi_{1s} \phi_{2s} \phi_{2s}^*=0
\label{4}
\end{eqnarray} 
\begin{eqnarray}
& & i\frac{\partial \phi_{2s}}{\partial z} + i\beta'_{s}\frac{\partial \phi_{2s}}{\partial t}
\nonumber\\
& &  + \overbrace{\frac{2}{3} \left( C_a(z)  + \frac{1}{3} C_b(z) \right) \phi_{2p}^2 \phi_{2i}^*e^{-i \Delta k z}}^{X_3} + 
 \frac{2}{3} \left( 2 + \frac{2}{3} C_a(z) \right) \phi_{2i} \phi_{2s} \phi_{2i}^* 
\nonumber\\
& & + \frac{2}{3} \left( 2 + \frac{2}{3} C_a(z) \right) \phi_{2p} \phi_{2s} \phi_{2p}^* + \frac{8}{9} \phi_{2s}^2 \phi_{2s}^* + 
 \frac{2}{3} \left( 2 - \frac{2}{3} C_a(z) \right) \phi_{2s} \phi_{1i} \phi_{1i}^* 
\nonumber\\
& & + \overbrace{\frac{2}{3} \left( C_a(z)  + \frac{1}{3} C_b(z) \right) \phi_{2p} \phi_{1p} \phi_{1i}^*e^{-i \Delta k z}}^{X_4} 
+ \frac{8}{9} C_a(z)  \phi_{2i} \phi_{1s} \phi_{1i}^* 
\nonumber\\
& & + \frac{2}{3} \left( 2 - \frac{2}{3} C_a(z) \right) \phi_{2s} \phi_{1p} \phi_{1p}^* + 
 \frac{8}{9} C_a(z)  \phi_{2p} \phi_{1s} \phi_{1p}^* + 
 \frac{8}{9} \phi_{2s} \phi_{1s} \phi_{1s}^*=0
\label{5}
\end{eqnarray}
Here $C_a(z)=\exp [-(8/3)(\Delta^{(-)})^2L_c z]$ and $C_b(z)=\exp [-(1/3)(\Delta^{(-)})^2L_c z]$, where 
$\Delta^{(-)}=\Delta\beta (\omega_p)-\Delta\beta (\omega_{i})$. The equation for $\phi_{1i}$ is obtained from Eq.(\ref{4}) by exchanging the labels $i$ and 
$s$; the equations for $\phi_{2p}$ and $\phi_{2i}$ are obtained from the equations for $\phi_{1p}$ and $\phi_{1i}$, respectively, by exchanging the labels 
$1$ and $2$. The polarization components $\phi_{1f}$ and $\phi_{2f}$ are obtained from the original components $u_{xf}$ and $u_{yf}$ by means of a 
unitary tranformation of the reference frame, see Appendices  A and B for details.

When the value of $\Delta^{(-)}$ is close to zero, the $z$-dependent coefficients $C_a(z)$ and $C_b(z)$ are both equal to unity, and we restore the 
model equations derived in Ref.~\cite{mckinstrie2004} starting from the Manakov equation.  This limit corresponds to vanishing PMD, and it is quite 
natural to call it as Manakov limit. In this limit the conversion of the pump energy into the signal is maximally efficient, and therefore parametric amplifiers 
should be designed in such a way that the FWM diffusion length $L_{fwm}\equiv  [(8/3)(\Delta^{(-)})^2L_c]^{-1}$ is much longer than
the fiber length $L$. To the best of our knowledge, our theory for the first time analytically predicts the length scale ($L_{fwm}$) of degradation of the 
process of parameteric amplification in telecom fibers. Strictly speaking, our theory is valid in two limits: $L\ll L_{fwm}$ and $L\gg L_{fwm}$. In the
opposite limit (the limit of large PMD, which we call diffusion limit) where $L\gg L_{fwm}$, the FWM process  is totally suppressed. Therefore this regime 
is not interesting from the viewpoint of frequency conversion. Most likely, the intermediate case of $L\sim L_{fwm}$ can be adequately treated only 
numerically, however we believe that the exponential decay of the nonlinear coefficients provides a qualitatively correct description of the rate of 
degradation. 

Another length scale ($L_d\equiv  [(1/3)(\Delta^{(-)})^2L_c]^{-1}$) appears in Eqs.(\ref{3}-\ref{5}), which is called the PMD diffusion length. This length is a 
characteristic which also enters the theory of two-beam nonlinear interactions. This length scale was introduced by Lin and Agrawal in 
Ref.~\cite{Agrawal_Raman} in the context of fiber-optic Raman amplifiers, and it was identified as the typical length at which the mutual orientation
of the states of polarization of the pump and signal beams is scrambled as a result of PMD. It is quite remarkable that this very same length scale not only 
characterizes the ``polarization memory" of Raman interactions, but it also characterizes the degradation of cross-polarization modulation (XPolM) 
mediated Kerr-interactions of the two beams, as it was shown in Ref.~\cite{Springer}.

The comparison of the two length scales, $L_{fwm}$ and $L_d$, shows that polarization sensitive frequency conversion is a more demanding process 
than polarization sensitive Raman amplification and XPolM-induced polarization attraction in telecom fibers, in the sense that its degradation is 
characterized by an eight times faster spatial degradation rate. In order to overcome this detrimental degradation process an experimentalist needs to 
select an ultra-low PMD highly nonlinear fiber. As the degradation rate depends quadratically on the frequency difference between signal and pump, the 
effect of PMD can be also viewed as setting an upper limit to the bandwidth of the simultaneous parametric amplification and repolarization process.

%===================================
\section{Polarization attraction: analytical results}
In this section we shall apply Eqs.(\ref{3}-\ref{5}) to describe the effect of polarization attraction of a signal or idler wave towards the SOP of a 
copropagating pump beam by means of FWM in a randomly birefringent telecom fiber. We will limit our analysis here to the small signal case, i.e., we 
make the undepleted pump approximation. As we shall see, this approximation permits us to obtain relatively simple analytical results for the effective 
bandwidth and gain of the polarization attraction process. From Eq.(\ref{3}) one obtains the two polarization components of the pump amplitude as
\begin{eqnarray}
\phi_{1p}=\sqrt{P}\exp(i\theta_{1p0} + i\gamma(8/9)P_{tot}z )
\nonumber\\
\phi_{2p}=\sqrt{Q}\exp(i\theta_{2p0} + i\gamma(8/9)P_{tot}z )
\label{pumpSolutions}
\end{eqnarray}
\noindent where $P=\vert\phi_{1p}(0)\vert^2$, $Q=\vert\phi_{2p}(0)\vert^2$, $\sqrt{P}\exp(\theta_{1p0})$ and $\sqrt{Q}\exp(\theta_{2p0})$ are the
input pump amplitudes in the fiber and $P_{tot} = P + Q$ is the conserved total pump power. Labels $X_{(1,2,3,4)}$ in Eqs.(\ref{3}-\ref{5}) indicate the 
four different FWM processes leading to sideband gain through the conversion of two pump photons in two sidebands photons. For vanishing PMD 
(Manakov limit) $\Delta^{(-)}\cong 0$, $C_a(z)\cong C_b(z)\cong 1$: peak sideband gain is obtained at a frequency detuning $\Delta \omega_{p1}$ 
such that the dispersive mismatch is compensated by the pump-induced nonlinear phase shift, i.e., 
$\beta_2\Delta\omega_{p1}^2+\gamma(16/9)P_{tot}=0$. In the absence of higher-order dispersion, this condition can only be reached in the anomalous 
dispersion regime (i.e., with $\beta_2 \le 0$).  In the opposite case of large PMD (diffusion limit) one has $C_a(z)\cong C_b(z)\cong 0$, so that the FWM 
terms are effectively suppressed. Yet, approaching the diffusion limit is equivalent to reducing the effective pump power to zero, which correspondingly 
leads to a peak gain for $\Delta \omega_{p2}\cong 0$. Let us consider now the intermediate case of $L\sim L_{fwm}$, where peak gain is observed at an 
intermediate sideband detuning $\Delta \omega_{p3}$ where $\Delta \omega_{p2}\leq\Delta \omega_{p3}\leq\Delta \omega_{p1}$. 
In order to quantify the sideband gain and evaluate their SOP relative to the pump we need to solve Eqs.(\ref{3}-\ref{5}). Let us apply the change of 
variables $\phi_{1,2(i,s)}=\tilde{\phi}_{1,2(i,s)}\exp(-ivz/2+i\theta_{P,Q})$, where $v=\beta_2 \Delta\omega^2-(16/9)\gamma P_{tot}$. By linearizing 
Eqs.(\ref{4},\ref{5}) for the sidebands, one obtains
\begin{eqnarray}
\frac{\partial\vec{\tilde{\phi}}}{\partial z}=i\frac{8}{9}M(z)\vec{\tilde{\phi}}
\nonumber
\label{equSidebands}
\end{eqnarray}
\noindent where $\vec{\tilde{\phi}}=[\tilde{\phi}_{1i},\tilde{\phi}_{1s}^*,\tilde{\phi}_{2i},\tilde{\phi}_{2s}^*]^T$, and 
M(z)=\\
\\
$\left[\begin{smallmatrix}
F_A(z) P +F_B(z) Q+v/2 & F_C(z) P & F_D(z)\sqrt{PQ} & F_C(z)\sqrt{PQ} \\
-F_B(z) P & -F_A(z) P -F_B(z) Q-v/2   & -F_C(z)\sqrt{PQ} & -F_D(z)\sqrt{PQ} \\
F_D(z)\sqrt{PQ} & F_C(z)\sqrt{PQ} & F_B(z) P +F_A(z) Q+v/2  & F_C(z) Q \\
-F_C(z)\sqrt{PQ} & -F_D(z)\sqrt{PQ} & -F_C(z) Q & -F_B(z) P -F_A(z) Q-v/2  \\
\end{smallmatrix}\right]$
\noindent with $F_A(z)=4\gamma/3(1+C_a(z)/3)$, $F_B(z)=4\gamma/3(1-C_a(z)/3)$, 
$F_C(z)=2\gamma /3(C_a(z)+C_b(z)/3)$; $F_D(z)=8\gamma C_a(z)/9$.
The solution of Eqs.(\ref{equSidebands}) may be written as 
\begin{eqnarray}
\vec{\tilde{\phi}}(z=L)=\exp(\Omega(L))\vec{\tilde{\phi}}(z=0)
\nonumber
\label{equSidebandssol}
\end{eqnarray}
where $\Omega (z)$ is constructed from $M(z)$ as a Magnus series expansion~\cite{Magnus}. Whenever the $z$-dependent coefficients 
$F_{(A,B,C,D)}(z)$ are slowly-varying over $L$ (i.e., $L_{fwm}\ge L$) we may truncate the expansion after the first term
\begin{equation}
\Omega (L)\cong\Omega_1(L)=\int_{z=0}^L M(z)dz\equiv\bar{M} 
\end{equation}
so that we simply replace $F_{(A,B,C,D)}(z)$ with their average values
\begin{equation}
\bar{F}_{(A,B,C,D)}=\frac{1}{L}\int_{y=0}^L F_{(A,B,C,D)}(z)dz
\label{meanApprox}
\end{equation}
which can be analytically calculated since $\bar{C}_a=k_a^{-1}( 1 - \exp(-k_a L) )$ and $\bar{C}_b=k_b^{-1}( 1 - \exp(-k_b L) )$, where 
$k_a=(1/3)(\Delta^{(-)})^2L_c$ and $k_b=(8/3)(\Delta^{(-)})^2L_c$. In the anomalous dispersion regime and for sideband frequency detunings 
$\Delta\omega$ below a certain cut-off value $\Delta\omega_c$, $\bar{M}$ has an eigenvalue with positive imaginary part, leading to the effective (or 
average) sideband gain coefficient $g_e$
\begin{eqnarray}
g_e^2= \frac{4}{90}\gamma^2P_{tot}^2\Bigg(-4+5\bar{C}_a^2 + \bar{C}_b^2 -8\bar{C}_a + 6\bar{C}_a\bar{C}_b\Bigg)
\nonumber\\ 
-\frac{1}{4}\beta_2^2\Delta\omega^4 
- \frac{4}{9}\beta_2\Bigg(1+\bar{C}_a\Bigg)\gamma P_{tot}\Delta\omega^2
\label{gain}
\end{eqnarray} 
\noindent From Eq.(\ref{gain}) we obtain the cut-off frequency $\Delta\omega_c$ of the gain band, 
the peak frequency detuning $\Delta \omega_{p4}$ and effective gain $g_{e,peak}$ as
\begin{flushleft}
\begin{eqnarray}
&& \Delta\omega_c^2 = 4c\Bigg(\frac{6\gamma L_c^{-1} P_{tot}}{27\vert\beta_2\vert c^2 L_c^{-1}+82\Delta n^2\gamma P_{tot} L}\Bigg),
\nonumber\\
&& \Delta\omega_{p4} = \Delta\omega_c/\sqrt{2},
\nonumber\\
&& g_{e,peak}^2=\frac{8}{3} \Bigg(\frac{\gamma^2 P_{tot}^2 (3 |\beta_2| c^2 L_c^{-1} - 2 \Delta n^2 \gamma P_{tot} L)}
{27 \beta_2 c^2 L_c^{-1} + 82 \Delta n^2 \gamma P_{tot} L}\Bigg)
\label{peaks}
\end{eqnarray}
\end{flushleft}
\noindent where $c$ is the speed of light in vacuum. Let us briefly discuss the role of the different physical parameters in determining the sideband gain 
and its bandwidth. First of all, increasing the birefringence strength $\Delta n$ or the fiber length $L$ reduces the peak gain coefficient as well as the 
optimal sideband detuning. In order to study the polarization properties of the sidebands we need to consider the eigenvectors of $\bar{M}$. 
For any frequency detuning $\Delta\omega$ within the gain band, let us denote by $\vec{p}$ the eigenvector of $\bar{M}$ which grows as 
$\exp (g_ez)$. After a relatively short distance into the fiber, we may well approximate the sideband fields as 
$\vec{\tilde\phi}\approx C \vec{p}\exp (g_ez)$, where $C$ is the projection or scalar product (which we suppose nonzero for simplicity) of the input 
sidebands polarization vector $\vec{\tilde{\phi}}(z=0)$ on $\vec{p}$. The components of $\vec{p}$ are such that
$\vec{p}[1]/\vec{p}[3]=\vec{p}[2]/\vec{p}[4]=\sqrt{P/Q}$. Idler amplitudes $\tilde{\phi_{1i}}$ and $\tilde{\phi_{2i}}$ correspond to the first and third 
components of $\vec{\tilde{\phi}}$, respectively. Thus their ratio can be expressed as $\tilde{\phi_{1i}}/\tilde{\phi_{2i}}= \vec{p}[1]/\vec{p}[3]=\sqrt{P/Q}$. 
Since $\phi_{1i}/\phi_{2i}=(\tilde{\phi_{1i}}/\tilde{\phi_{2i}})e^{i\theta_P-i\theta_Q}$, we obtain that
\begin{eqnarray}
\frac{\phi_{1i}}{\phi_{2i}}=\frac{\sqrt{P}}{\sqrt{Q}}e^{i\theta_P-i\theta_Q}= \frac{\phi_{1p}(z=0)}{\phi_{2p}(z=0)}
\label{attraction}
\end{eqnarray}
A similar treatment can be developed for the signal amplitudes too,
which proves the polarization attraction of both the signal and the idler to the input polarization of the pump.\\
In practice, since for a given sideband frequency detuning 
the effective gain coefficient $g_e$ decreases as the fiber length $L$ grows larger, the corresponding strength of polarization attraction will be reduced 
whenever the fiber length approaches $L_{fwm}$. As a matter of fact, in the diffusion limit $g_e=0$ and FWM-induced polarization attraction is no longer 
observed. In the next section we will provide a quantitative description of the fiber length dependence of the polarization attraction efficiency.  

%==============================
\section{Polarization attraction: examples}
Let us study the efficiency of polarization attraction as a function of fiber length $L$, hence of PMD. Consider a fiber with the nonlinear coefficient 
$\gamma= 11.9$~W$^{-1}$~km$^{-1}$, dispersion $\beta_2=-0.5$~ps$^2$~km$^{-1}$, and PMD correlation length $L_c=10$~m. The chosen parameters
are typical for highly nonlinear optical fibers. As well known, the SOP of each interacting wave may be represented by means of its corresponding 
unitary dimensionless Stokes vector as $ \vec{S}_j=[\,S_{1j}= S_{0j}^{-1}(\phi_{1j}^*\phi_{2j}+\phi_{1j}\phi_{2j}^*)$,
$S_{2j}= S_{0j}^{-1}(i \phi_{1j}^*\phi_{2j}-i \phi_{1j}\phi_{2j}^*)$, $S_{3j}= S_{0j}^{-1}(\vert\phi_{1j}\vert^2-\vert\phi_{2j}\vert^2)\,]$ $(j=i,p,s)$, where 
$S_{0j}=\left(\vert\phi_{1j}\vert^2+\vert\phi_{2j}\vert^2\right)^{1/2}$. The input CW pump beam power is set to $P_{tot}=S_{0p}=1$~W, and its SOP is
defined by the Stokes vector $\vec{S}_p=[ \sqrt{0.5},\,\sqrt{0.4},\,\sqrt{0.1} ]$. We set the input signal power to $P_{s,in}=1$~mW, whereas the idler is 
zero at the fiber input, as in typical FWM experiments. We compared the numerical solution of Eq.~(\ref{equSidebands}) with the analytical solution of 
Eq.~(\ref{equSidebandssol}). As initial condition we employed a set of $10000$ input signal SOPs, whose corresponding Stokes vectors are uniformly 
distributed over the Poincar\'e sphere. Figures \ref{f1} and \ref{f2} illustrate the dependence on sideband detuning of the signal gain $g_s$ and its output 
DOP, respectively, for four different values of $\Delta n$ (namely, $\Delta n=0,\, 10^{-6},\,1.5\cdot 10^{-6}$ and $\Delta n=1\cdot 10^{-5}$), and the fiber 
length $L=300$~m. The signal gain was computed as $g_s=(2L)^{-1}\log [P_{s,out}/P_{s,in}]$, where $P_{s,out}$ is the output signal power. The output 
DOP was calculated as discussed in \cite{errata}. In Figs.~\ref{f1} and \ref{f2} the curves refer to numerical solutions, and the dots to analytical solutions: 
as can be seen, the first-order term of the Magnus expansion provides an excellent approximation of the exact solution. Figs.~\ref{f1} show that, as the 
birefringence $\Delta n$ (or PMD) strength grows larger, the signal gain $g_s$ is progressively degraded; at the same time, both the peak gain 
frequency detuning $\Delta\omega_3$ and the cut-off frequency $\Delta\omega_c$ shrink towards zero. In addition Figs.~\ref{f2} show that the signal DOP is
maximum for sideband frequencies close to peak gain values: however the peak DOP rapidly drops from unity as the PMD strength is increased (i.e., for 
$\Delta n\ge 10^{-6}$). It is interesting to point out that, in contrast with the case of the signal, the output DOP of the idler (not shown here) remains close to unity throughout 
the entire gain bandwidth. The increased attraction of the idler towards the pump is due to the fact that the idler grows from zero at the fiber input, hence 
its projection on the growing eigenvector $\vec{p}$ is much larger than for the signal.  

In the second example of Fig.~(\ref{f3}) we show the signal DOP as a function of the fiber length $L$, for four different values of the sideband frequency 
detuning $\Delta\nu=\Delta\omega /(2\pi)$ (i.e., $\Delta\nu =0.255$~THz, $0.350$~THz, $0.365$~THz, $0.380$~THz): here the PMD value is kept fixed to 
$\Delta n=1.5\cdot 10^{-6}$. As it can be seen, for $\Delta\nu =0.255$~THz (which corresponds to the peak gain value $\Delta\omega_3$) the DOP is
monotonically increasing with distance and it approaches the unit value for $L\ge 500$~m. On the other hand, Fig.~(\ref{f3}) shows that for other values
of the sideband detuning the output DOP exhibits a damped oscillating behavior and it converges to relatively low values after fiber lengths of the order 
of $1$~km. 
 
It is useful to visualize the effectiveness of polarization attraction by means of parametric gain or FWM by plotting on the Poincar\'e sphere the end points 
of the Stokes vectors corresponding to either the input or the output distributions of signal SOPs, corresponding to the results of Figures \ref{f1}-\ref{f3}. 
In Fig.~\ref{f4} we compare the distribution of input signal SOPs, which uniformly covers the sphere ( Fig.~\ref{f4}(a) ), to the output signal SOP distribution ( Fig.~\ref{f4}(b) ) from a fiber of length $L=500$~m with birefringence $\Delta n=1.5\cdot10^{-6}$: here the signal detuning is $\Delta\nu=0.255$~THz. These parameters 
correspond to the sideband detuning for peak signal gain (see the dotted curve in Fig.~\ref{f2}). As it shown by the solid curve in Fig.~\ref{f3},
the output DOP is as high as $0.97$, which means a nearly full attraction towards the input pump Stokes vector $\vec{S}_p$. On the other hand, in 
Fig.~\ref{f5} we show the output distributionof signal SOPs when the sideband detuning is increased up to $\Delta\nu=0.350$~THz. Fig.~\ref{f3} shows
that the output DOP is only $0.73$ in this case, which results in a relatively poor polarization attraction. It is important to point out that the polarization 
attraction (to the pump SOP) behavior which is illustrated in Figs.~\ref{f4} and \ref{f5} does not depend upon the specific input pump SOP which is 
selected: indeed, the strength of polarization attraction only depends on the pump power level.

%===============
\section{Conclusions}

In our study we proposed and analysed a novel type of nonlinear polarizer, exploiting the degenerate FWM process or parametric optical amplification in
a standard telecom fiber with randomly varying birefringence. In the FWM-polarizer the SOP of the amplified signal (or idler) beam is attracted to the SOP 
of the copropagating, fully polarized pump wave. We have derived the coupled wave equations that describe the propagation of the pump, the idler, and 
the signal in the presence or weak PMD. Our model substantially extends previous theory of FWM in optical fibers, since it may analytically describe the 
rate of degradation of FWM efficiency and polarization attraction for low-PMD fibers. Knowing the spatial rate of PMD-induced degradation permits the 
proper design of practical nonlinear polarizers based on optical parametric amplification in km long nonlinear optical fibers. Polarization attraction and 
control by parametric amplification in fibers is potentially applicable to frequency-conversion and phase sensitive amplification devices when combined 
with polarization-sensitive optical processing devices (e.g., an heterodyne receiver). In addition, codirectional parametric repolarizers based on low-PMD
telecom fibers may be used for compensating ultrafast input signal SOP fluctuations. Although FWM-based polarizers suffer from output RIN, however 
RIN suppression could be obtained when operating the amplifier in the depleted pump regime, as it occurs with Raman polarizers ~\cite{kozlovPTL}. 

\section{Acknowledgements}
This work was carried out with support from the Italian Ministry of Research and the University (MIUR) through the grant 2008MPSSNX.

\appendix
%================================================
\section{Appendix: Stochastic theory of parametric amplification}
Our goal is to convert the initial equations for the field (\ref{1}) and (\ref{2}) with stochastic coefficients into corresponding equations with deterministic 
coefficients. In other words, we need to find a way to average the initial equations over the ensemble of fibers, which represents all possible realizations 
of the random fiber birefringence with a given statistics. Since both initial and final equations are nonlinear, our procedure cannot be done exactly and it 
will require a number of approximations. Thus, the final equations will have a limited range of applicability.

We use the approach first introduced into the fiber optics theory by Wai and Menyuk in Ref.~\cite{Wai_Menyuk}. This approach was formulated for a 
single beam (or pulse), and lead to the derivation of the celebrated Manakov equation and its generalization in the form of the Manakov-PMD equation. 
An extension of this theory for the two-beam configuration was undertaken in Refs.~\cite{Kozlov2011,KozlovJLT}, and led to the formulation of the 
theoretical basis of XPolM-induced polarization attraction effect in telecom fibers and of Raman polarizers. Here, we need to extend this theory even 
further by fully taking into account the three interacting beams. Given that all these theories have very much in common, we shall omit many repetitions 
and where appropriate we simply refer to prior literature for more details. 

We start with the transformation of field vectors from the laboratory $(x,\, y)$ frame into the local reference frame $(1,\, 2)$, which is defined by the 
$z$-dependent orientation of the axis of birefringence: $\Psi_f=M(z)U_f$, where $M(z)$ is the $2\times 2$ rotation matrix defined in Eq.~(4) of 
Ref.~\cite{Kozlov2011}. Here $\Psi_f=(\psi_{1f},\, \psi_{2f})$. All terms except one in the field equations stay immune to this transformation. The only
change is the form of the birefringence matrix, which now becomes
\begin{equation}
\overline{\Delta B}(\omega_{f})=\left(
\begin{array}{cc}
\Delta\beta (\omega_{f}) & \mp\frac{i}{2}g_\theta\\
\pm\frac{i}{2}g_\theta &  -\Delta\beta (\omega_{f})
\end{array}
\right)\, .
\label{A1}
\end{equation}
The next transformation: $\Phi_{f}=T_f(z)\Psi_{f}$, is aimed at the decoupling of the linear portions of the field equations. This goal is reached if the
transformation matrix
\begin{equation}
T_p(z)=\left(
\begin{array}{cc}
a_1(z) & a_2(z) \\
-a_2^*(z) & a_1^*(z)
\end{array}
\right)
\label{A2}
\end{equation}
obeys the following equation
\begin{equation}
i\frac{\partial T_p}{\partial z}+\overline{\Delta B}_p\cdot T_p=0\, .
\label{A3}
\end{equation} 
Matrices $T_s$ and $T_i$ are defined in a similar way, with $b_{1,2}$ and $c_{1,2}$ elements used instead of $a_{1,2}$. The unitarity of this 
transformation is preserved by requiring that 
$\vert a_1(z)\vert^2+\vert a_2(z)\vert^2=\vert b_1(z)\vert^2+\vert b_2(z)\vert^2=\vert c_1(z)\vert^2+\vert c_2(z)\vert^2=1$. 
Initial conditions for the elements of the $T_f(z)$ matrices are to be determined from the requirement that $\Phi_f=\Psi_f$ at $z=0$. Thus, 
$a_1(0)=b_1(0)=c_1(0)=1$ and $a_2(0)=b_2(0)=c_2(0)=0$.

The transformation associated with the $T_f(z)$ matrix brings the equations for three fields in the form
\begin{equation}
i\frac{\partial \Phi_f}{\partial z} + \gamma (N_{spm}+N_{xpm}+N_{ex})_f=0\, .
\label{A4}
\end{equation}
As expected, in this reference frame the fields are coupled by nonlinearity only through three types of cubic terms: SPolM terms $N_{spm}$, XPolM 
terms $N_{xpm}$, and energy exchange terms $N_{ex}$. The number of these nonlinear terms is very large, and we do not provide here their detailed 
structure. Instead, we refer to Eqs.~(9)-(12) in Ref.~\cite{Kozlov2011} where the SPolM and XPolM nonlinear terms are written down explicitely. In our 
present theory we have all these terms as well, and in addition get energy-exchange terms in the form of cubic products involving three different fields.

Coefficients prior to these terms are some self- and cross- fourth-order polynomials composed of $a_{1,2}(z)$, $b_{1,2}(z)$, $c_{1,2}(z)$ and their 
complex conjugates. It is convenient to work with quadratic coefficients $u_m$ and $u_m^*$ ($m=1\div 30$). Coefficients $u_m$ with $m=1\div 14$ are 
identical to those introduced immediately below Eq.~(12) in Ref.~\cite{Kozlov2011}. They are divided into self-terms:
$u_1=\vert a_1\vert^2-\vert a_2\vert^2$,
$u_2=-(a_1a_2+a_1^*a_2^*)$,
$u_3=i(a_1a_2-a_1^*a_2^*)$,
$u_4=2a_1a_2^*$,
$u_5=a_1^2-{a_2^*}^2$,
$u_6=-i(a_1^2+{a_2^*}^2)$,
and cross-terms:
$u_7=a_1^*b_1-a_2b_2^*$,
$u_8=-(b_1a_2+b_2^*a_1^*)$,
$u_9=i(b_1a_2-a_1^*b_2^*)$,
$u_{10}=-i(a_1^*b_1+a_2b_2^*)$,
$u_{11}=a_1b_2^*+b_1a_2^*$,
$u_{12}=a_1b_1-a_2^*b_2^*$,
$u_{13}=-i(a_1b_1+a_2^*b_2^*)$,
$u_{14}=i(a_1b_2^*-a_2^*b_1)$.
The three-beam theory additionally brings $16$ new coefficients. Coefficients $u_m$ with $m=15\div 22$ are the same as $u_m$ with $m=7\div 14$ but 
with $b_{1,2}$ replaced with $c_{1,2}$. Coefficients $u_m$ with $m=23\div 30$ are the same as $u_m$ with $m=7\div 14$ where $a_{1,2}$ is replaced 
with $b_{1,2}$, and simultaneously $b_{1,2}$ is replaced with $c_{1,2}$.

Nonlinear coefficients in Eq.~(\ref{A3}) are products of the type $u_mu_n$ or $u_mu_n^*$. They are $z$-dependent random coefficients, because they 
depend on the stochastic variable $g_\theta (z)$. We need to find average values of all nonlinear terms, which are of the form, for instance, 
$u_9^2\phi_{1s}\phi_{2p}\phi_{2s}^*$. This is the place where the most important approximation comes into play. We assume that the following 
factorization is valid: $\langle u_9^2\phi_{1s}\phi_{2p}\phi_{2s}^*\rangle \approx\langle u_9^2\rangle\,\langle\phi_{1s}\phi_{2p}\phi_{2s}^*\rangle$. 
This factorization is justified whenever the spatial evolution of the fields is much slower than the spatial evolution of the nonlinear coefficients, or vice 
versa, whenever the spatial evolution of the fields is much faster than the spatial evolution of the nonlinear coefficients. In the context of parametric 
amplification and in the absence of group-velocity dispersion, the nonlinear evolution of the fields scales with the nonlinear length 
$L_{NL}=\gamma P_p$, where $P_p$ is the pump power. In its turn, the $z$-dependence of $\langle u_mu_n\rangle$ or $\langle u_mu_n^*\rangle$ is 
governed by two different length scales. On the one hand we have the relatively short spatial scales which are associated with the correlation length 
$L_c$ and the beat length $L_B$, both of which are typically less than $100$~m. On the other hand we have the relatively long spatial scales which are 
associated with the PMD diffusion length $L_d$ and the FWM diffusion length $L_{fwm}$. For practically interesting situations we need to provide the 
following hierarchy of scales: $L_c,\, L_B\ll L,\, L_{NL}\ll L_{d},\, L_{fwm}$. In this range, the factorization approximation is well justified: with this limitation 
in mind, we may proceed further.

In order to find averages of the type $u_m^2$ and $\vert u_m\vert^2$, it is convenient to group coefficients as
$G_1=\{ u_1,\, u_2,\, u_3\}$, $G_2=\{ u_4,\, u_5,\, u_6\}$, $G_3=\{ u_7,\, u_8,\, u_9,\, u_{10}\}$, $G_4=\{ u_{11},\, u_{12},\, u_{13},\, u_{14}\}$,
$G_5=\{ u_{15},\, u_{16},\, u_{17},\, u_{18}\}$, $G_6=\{u_{19},\, u_{20},\, u_{21},\, u_{22}\}$, $G_7=\{u_{23},\, u_{24},\, u_{25},\, u_{26}\}$,
and $G_8=\{ u_{27},\, u_{28},\, u_{29},\, u_{30}\}$. For each group we were able to formulate a closed system of linear first-order differential equations by 
using Eq.~(\ref{A3}). For an example of such a system, we may refer to Eq.~(13) in Ref.~\cite{Kozlov2011}.  

Next we need to know the average values of quadratic forms composed by these coefficients. They can be found from the solutions to the equations of 
motion for the average of the generic function $F$. For instance, for $F(u_1,u_2,u_3,\theta )$ we need to solve the equation $\partial_z\langle F\rangle 
=\langle G(F)\rangle$. The generator $G$ is to be constructed by a procedure described in the Appendix of Ref.~\cite{Wai_Menyuk}. For a specific 
example of $G$, we may refer to Eqs.~(14) and (20) in Ref.~\cite{Kozlov2011}. Note also that the average over different realizations of the fiber 
birefringence can be replaced by a spatial average as
\begin{equation}
\langle f\rangle =\lim_{z\to\infty} \frac{1}{z}\int_0^z \! dz^\prime\, f(z^\prime )\, ,
\label{A5}
\end{equation}
by assuming that the ergodicity hypothesis is valid.

With this procedure at hand, we are able to find the mean values of $u_m^2$, with $m=9,\, 10,\, 13,\, 14,\, 17,\, 18,\, 21,\, 22,\, 25,\, 26,\, 29,\, 30$ by 
solving the equation $\partial_z V_A=M_AV_A$ for the vector $V_A= (\langle S_1^2\rangle ,\, \langle S_2^2\rangle ,\,  \langle S_3^2\rangle ,\,  \langle 
S_4^2\rangle ,\, \langle S_2S_3\rangle ,\, \langle S_1S_4\rangle )^T$, where  $\{S_1,\, S_2,\, S_3,\, S_4\}$ is any of the groups $G_i$ with
$i=4\div 8$, and with the matrix $M_A$ given by
\begin{equation}
\begin{pmatrix} 
-2L_c^{-1} & 2L_c^{-1} & 0& 0 & 0 & 2\Delta^{-}  \\ 
2L_c^{-1} & -2L_c^{-1} & 0 & 0 & -2\Delta^{+} & 0 \\ 0 & 0 & 0& 0&  2\Delta^{(+)} & 0 \\ 
0 & 0 & 0& 0& 0& -2\Delta^{(-)}\\ 
0 & \Delta^{(+)} & -\Delta^{(+)}& 0& -L_c^{-1}& 0 \\ 
-\Delta^{(-)} & 0 & 0& \Delta^{(-)}& 0& -L_c^{-1} 
\end{pmatrix} 
\label{A6}
\end{equation}
Here $\Delta^{(+)}=\Delta\beta (\omega_p)+\Delta\beta (\omega_{i})$ and $\Delta^{(-)}=\Delta\beta (\omega_p)-\Delta\beta (\omega_{i})$. It is a 
straightforward calculation to get an estimate $\Delta^{(-)}/\Delta^{(+)}\sim \Delta\omega /\omega_p$, where $\Delta\omega 
=\omega_p-\omega_i=\omega_s-\omega_p$. For typical fiber parameters, the evolution associated with $\Delta^{(+)}$ is very fast, while $\Delta^{(-)}$ 
defines a much slower spatial scale. Setting $\Delta^{(-)}$ to zero defines the Manakov limit, and brings us back to the formulation of coupled 
wave equations with constant in $z$ nonlinear coefficients. The difference of $\Delta^{(-)}$ from zero means the inclusion of effects caused by the PMD. In this case we 
are dealing with $z$-dependent nonlinear coefficients.  

Next we calculate the averages of the type $\vert u_m\vert^2$, with $m=9,\, 10,\, 13,\, 14,\, 17,\, 18,\, 21,\, 22,\, 25,\, 26,\, 29,\, 30$. Thereto we formulate
the equation of motion $\partial_z V_B=M_BV_B$ for the vector $V_B=(\langle \vert S_1\vert^2\rangle ,\,  \langle\vert S_2\vert^2\rangle ,\,  
\langle\vert S_3\vert^2\rangle ,\,  \langle\vert S_4\vert^2\rangle ,\,  \langle S_2S_3^*\rangle ,\,  \langle S_2^*S_3\rangle ,\, 
\langle S_1S_4^*\rangle ,\,  \langle S_1^*S_4\rangle )^T$ where $\{ S_1,\, S_2,\, S_3,\, S_4\}$ is any of the groups $G_i$, with $i=4\div  8$, and where the
matrix $M_B$ reads as
\begin{equation}
\begin{pmatrix} 
-2L_c^{-1} & 2L_c^{-1} & 0& 0 & 0 & 0 & \Delta^{(-)}  & \Delta^{(-)} \\ 
2L_c^{-1} & -2L_c^{-1} & 0 & 0 & -\Delta^{+} & -\Delta^{+}& 0& 0 \\ 
0 & 0 & 0 & 0 &  \Delta^{(+)} & \Delta^{(+)}& 0 & 0 \\ 
0 & 0 & 0& 0 & 0&  0 & -\Delta^{(-)}& -\Delta^{(-)} \\  
0 & \Delta^{(+)} & -\Delta^{(+)}& 0& -L_c^{-1}& 0 & 0 & 0  \\ 
0 & \Delta^{(+)} & -\Delta^{(+)}& 0& 0& -L_c^{-1}& 0 & 0  \\  
-\Delta^{(-)} & 0 & 0& \Delta^{(-)}& 0 & 0 & -L_c^{-1} & 0 \\  
-\Delta^{(-)} & 0 & 0& \Delta^{(-)} & 0 & 0 & 0 & -L_c^{-1}
\end{pmatrix} 
\label{A7}
\end{equation}

Note that initial conditions for the averages of the type $\langle u_m(z)u_n(z)\rangle$ and $\langle u_m(z)u_n^*(z)\rangle$ can be found from the initial 
conditions for the coefficients $a_{1,2}$, $b_{1,2}$, and $c_{1,2}$, and by observing that $\langle u_m(0)u_n(0)\rangle =u_m(0)u_n(0)$. Thus we find 
$u_{1,5,7,12,15,20,25,28}(0)=1$ and $u_{6,10,13,17,21,26,29}(0)=-i$, while the remaining coefficients are all zero. 

Next, we turn to cross-terms like $\langle u_mu_n\rangle$ with $m\ne n$. Many of these terms are zero, mainly because of the imposed zero initial 
conditions.  Nonzero coefficients are $\langle u_{14}u_{22}\rangle$, $\langle u_{14}u_{22}^*\rangle$, $\langle u_{10}u_{18}\rangle$, 
$\langle u_{10}u_{18}^*\rangle$, $\langle u_6u_{29}\rangle$, $\langle u_6u_{29}^*\rangle$, $\langle u_3u_{25}\rangle$, and 
$\langle u_3u_{25}^*\rangle$.  The first four of these coefficients can be found by solving the equation $\partial_z V_{B}=M_BV_{B}$ with the matrix 
$M_B$ defined as in Eq.~(\ref{A7}), and where the vector $V_{B}$ is identified with any of the following vectors: 
$(\langle u_{11}u_{19}\rangle ,\, \langle u_{12}u_{20}\rangle ,\, \langle u_{13}u_{21}\rangle ,\, \langle u_{14}u_{22}\rangle ,\, \langle u_{13}u_{20}\rangle ,\,
 \langle u_{12}u_{21}\rangle ,\, \langle u_{14}u_{19}\rangle ,\, \langle u_{11}u_{22}\rangle )^T$, 
$(\langle u_{11}u_{19}^*\rangle ,\, \langle u_{12}u_{20}^*\rangle ,\, \langle u_{13}u_{21}^*\rangle ,\, \langle u_{14}u_{22}^*\rangle ,\, 
\langle u_{13}u_{20}^*\rangle ,\, \langle u_{12}u_{21}^*\rangle ,\, \langle u_{14}u_{19}^*\rangle ,\, \langle u_{11}u_{22}^*\rangle )^T$,
$(\langle u_{7}u_{15}\rangle ,\, \langle u_{8}u_{16}\rangle ,\, \langle u_{9}u_{17}\rangle ,\, \langle u_{10}u_{18}\rangle ,\, 
\langle u_{9}u_{16}\rangle ,\, \langle u_{8}u_{17}\rangle ,\, \langle u_{10}u_{15}\rangle ,\, \langle u_{7}u_{18}\rangle )^T$,
$(\langle u_{7}u_{15}^*\rangle ,\, \langle u_{8}u_{16}^*\rangle ,\, \langle u_{9}u_{17}^*\rangle ,\, \langle u_{10}u_{18}^*\rangle ,\, 
\langle u_{9}u_{16}^*\rangle ,\, \langle u_{8}u_{17}^*\rangle ,\, \langle u_{10}u_{15}^*\rangle ,\, \langle u_{7}u_{18}^*\rangle )^T$.

In turn, the coefficients $\langle u_6u_{29}\rangle$, $\langle u_6u_{29}^*\rangle$, $\langle u_3u_{25}\rangle$, and 
$\langle u_3u_{25}^*\rangle$ can be found from the equation $\partial_z V_C=M_CV_C$ with the matrix $M_C$ defined as
\begin{equation}
\begin{pmatrix} 
0 & 0 & \Delta^{(+)}& \Delta^{(+)} & 0 & 0  \\ 
-\Delta^{(+)} & -L_c^{-1} & 0& \Delta^{(+)} & 0 & 0 \\ 
-\Delta^{(+)} & 0 & -L_c^{-1}& \Delta^{(+)}&  0 & 0 \\ 
0 & -\Delta^{(+)} & -\Delta^{(+)}& -2L_c^{-1}& 2L_c^{-1} & 0\\ 
0 & 0 & 0& 0& 2L_c^{-1}& -2L_c^{-1} \\ 
0 & 0 & 0 & 0& -\Delta^{(-)}& -L_c^{-1}
\end{pmatrix} 
\label{A8}
\end{equation}
when we associate the vector $V_C$ with any of the following vectors:
$(\langle u_{6}u_{29}\rangle ,\,\langle u_{5}u_{29}\rangle ,\,\langle u_{6}u_{28}\rangle ,\,\langle u_{5}u_{28}\rangle ,\,
\langle u_{4}u_{27}\rangle ,\,\langle u_{4}u_{30}\rangle )^T$, 
$(\langle u_{6}u_{29}^*\rangle ,\, \langle u_{5}u_{29}^*\rangle ,\, \langle u_{6}u_{28}^*\rangle ,\, \langle u_{5}u_{28}^*\rangle ,\, 
\langle u_{4}u_{27}^*\rangle ,\, \langle u_{4}u_{30}^*\rangle )^T$,
$(\langle u_{3}u_{25}\rangle ,\, \langle u_{3}u_{24}\rangle ,\langle u_{2}u_{25}\rangle ,\, \langle u_{2}u_{24}\rangle ,\, 
\langle u_{1}u_{23}\rangle ,\, \langle u_{1}u_{26}\rangle )^T$, and 
$(\langle u_{3}u_{25}^*\rangle ,\, \langle u_{3}u_{24}^*\rangle ,\, \langle u_{2}u_{25}^*\rangle ,\, \langle u_{2}u_{24}^*\rangle ,\,
 \langle u_{1}u_{23}^*\rangle ,\, \langle u_{1}u_{26}^*\rangle )^T$.

%==================================================
\section{Appendix: Analytic estimation of the nonlinear coefficients}
In this Appendix we look for approximate analytical solutions to the linear systems of equations for the vectors $V_A$, $V_B$ and $V_C$. This task is 
equivalent to finding the eigenvalues and eigenvectors of the matrices $M_{A,B,C}$. Additionally, we need to find the decomposition of the intial vectors
$V_{A,B,C}(0)$ in the basis of the corresponding eigenvectors. In this way we may determine the $z$-dependence of the nonlinear coefficients.

We shall give a detailed analysis for the $M_A$ matrix, and sketch only briefly the results for the other matrices. We develop a perturbative approach, by 
assuming that $\Delta^{(-)}$ is much smaller than $\Delta^{(+)}$ and $L_c^{-1}$. First, setting $\Delta^{(-)}$ to zero we get a much simpler matrix 
$\tilde{M}_A$. The difference $\Delta{M}_A=M_A-\tilde{M}_A$ is therefore a small correction. The matrix $\tilde{M}_A$ has a doubly degenerate 
eigenvalue $\tilde\lambda_A=0$ and two corresponding eigenvectors $\tilde{e}_{A1}=(0,0,0,1,0,0)^T$ and $\tilde{e}_{A2}=(1,1,1,0,0,0)^T$. The other 
eigenvalues of $\tilde{M}_A$ all have relatively large negative real parts, in the sense that the corresponding eigenvectors vanish with distance
very quickly. The spatial scale of this decay is determined by the correlation length $L_c$ and the beat length $L_B$, both of which are typically less than 
$100$~m. So, the characteristic decay rate is estimated as $L_{transient}\sim 100$~m. After the transient decay is over we can write the solution of 
$\partial_z \tilde{V}_A=\tilde{M}_A\tilde{V}_A$ as $\tilde{V}_A(z)=(\tilde{C}_1\tilde{e}_{A1} + \tilde{C}_2\tilde{e}_{A2})\exp (\tilde{\lambda}_A z)= 
\tilde{C}_1\tilde{e}_{A1} + \tilde{C}_2\tilde{e}_{A2}$, where $\tilde{C_1}=(V_A(0)\cdot \tilde{e}_{A1}^*)/(\tilde{e}_{A1}\cdot \tilde{e}_{A1}^*)$ and 
$\tilde{C}_2=(V_A(0)\cdot \tilde{e}_{A2}^*)/(\tilde{e}_{A2}\cdot \tilde{e}_{A2}^*)$, thanks to the orthogonality of the set of eigenvectors of $\tilde{M}_A$.

When $\Delta^{(-)}$ is different from zero, the degeneracy is lifted and the doubly degenerate eigenvalue $\tilde\lambda_A$ split into two different 
eigenvalues $\lambda_{A1}=0$ and $\lambda_{A2} \neq 0$. Let us find $\lambda_{A2}$ by way of developing the perturbative analysis. First we find the 
eigenvalue equation for the exact $M_A$ matrix. It is  $\hbox{det}({M_A-\lambda I}) = 32 [\Delta^{(-)}]^2 [\Delta{(+)}]^2 L_c^{-1}\lambda +  (16 
[\Delta^{(-)}]^2 [\Delta{(+)}]^2+12 [\Delta^{(-)}]^2  L_c^{-2}+12 [\Delta^{(+)}]^2  L_c^{-2}) \lambda^2 + (16 [\Delta^{(-)}]^2 L_c^{-1}+16 [\Delta{(+)}]^2 
L_c^{-1}+4 L_c^{-3})\lambda^3 + (4 [\Delta^{(-)}]^2+4 [\Delta{(+)}]^2+9 L_c^{-2})\lambda^4 + 6 L_c^{-1}\lambda^5 + \lambda ^6 = 0$, where $I$ is the unity 
matrix. Since $\Delta^{(-)}$ is small, we expect that the correction to the unperturbed zero eigenvalue $\tilde\lambda_A$ is also small. By keeping in
the eigenvalue equation terms no higher than second order in $\lambda$, we get after some simplifications the approximated solution
$\lambda_{A2}\approx -(8/3) (\Delta^{(-)})^2 L_c=-L_{fwm}^{-1}$. The eigenvector corresponding to eigenvalue $\lambda_{A1}$ ($\lambda_{A 2}$) is 
$e_{A1}=\tilde{e}_{A1}+\tilde{e}_{A2}$ ($e_{A2}$). The perturbed solution is 
$V_A(z)=C_1e_{A1}\exp (\lambda_{A1}z)+C_2e_{A2}\exp (\lambda_{A2}z)=C_1e_{A1}+C_2e_{A2}\exp (-z/L_{fwm})$. Here
$C_1=(V_A(0)\cdot e_{A1}^*)/(e_{A1}\cdot e_{A1}^*)$. The exact expression for $C_2e_{A2}$ is cumbersome, however under the condition 
$\Delta^{(-)}\ll \Delta^{(+)},\, L_c^{-1}$ we can use the equality of $\tilde{V}_A$ and $V_A$ in the limit of $\Delta^{(-)}\to 0$, and write 
$C_2e_{A2}=\tilde{C}_1\tilde{e}_{A1}+\tilde{C}_2\tilde{e}_{A2}-C_1e_{A1}$.

Now we can turn to evaluation of the nonlinear coefficients. Let us start with the averages $\langle u_9^2\rangle$ and $\langle u_{10}^2\rangle$. 
Coefficients $u_9$ and $u_{10}$ belong to the group of coefficients denoted earlier as $G_3$. For this group, the vector $V_A$ contains $\langle 
u_9^2\rangle$ and $\langle u_{10}^2\rangle$ as the third and the fourth element, respectively. The initial condition reads as $V_A(0)=(1,0,0,-1,0,i)^T$. 
Thus, for $L\ge L_{transient}$ we find $\langle u_9^2\rangle =(1/3)\exp (-z/L_{fwm})$ and $\langle u_{10}^2\rangle =-\exp (-z/L_{fwm})$. Similarly, we find
$\langle u_{17}^2\rangle = \langle u_{25}^2\rangle =\langle u_{9}^2\rangle$ and 
$\langle u_{18}^2\rangle = \langle u_{26}^2\rangle =\langle u_{10}^2\rangle$. With initial conditions $V_A(0)=(0,\, 1,\, -1,\, 0,\, -i,\, 0)^T$ we get
$\langle u_{13}^2\rangle =\langle u_{14}^2\rangle =\langle u_{21}^2\rangle =\langle u_{2}^2\rangle =\langle u_{29}^2\rangle =\langle u_{30}^2\rangle =0$.

The matrix $M_B$ can be considered similarly. Again, in the limit $\Delta^{(-)}\to 0$ this matrix possesses a doubly degenerate zero eigenvalue 
$\tilde\lambda_B$, while the other eigenvalues have large negative real parts, so that the corresponding eigenvectors vanish after a certain propagating 
distance, say $L_{transient}$. Whenever $\Delta^{(-)}$ is different from zero, the degeneracy is lifted and the doubly degenerate eigenvalue 
$\tilde\lambda_B$ is split into $\lambda_{B1}=0$ and $\lambda_{B2}=-L_{fwm}$. Thus, for $L\ge L_{transient}$ we find
$\langle\vert u_{9}\vert^2\rangle =\langle\vert u_{17}\vert^2\rangle = \langle\vert u_{25}^2\vert\rangle =(1/2)-(1/6)\exp(-z/L_{fwm})$;
$\langle\vert u_{10}\vert^2\rangle =\langle\vert u_{18}^2\vert\rangle =\langle\vert u_{26}\vert^2\rangle =(1/2)+(1/2)\exp(-z/L_{fwm})$;
$\langle\vert u_{13}\vert^2\rangle =\langle\vert u_{21}\vert^2\rangle = \langle\vert u_{29}^2\vert\rangle =(1/2)+(1/6)\exp(-z/L_{fwm})$;
$\langle\vert u_{14}\vert^2\rangle =\langle\vert u_{22}^2\vert\rangle = \langle\vert u_{30}\vert^2\rangle =(1/2)-(1/2)\exp(-z/L_{fwm})$.
By using the same eigenvectors and eigenvalues of matrix $M_B$ we find also that
$\langle u_{14}u_{22}\rangle =0$, $\langle u_{14}u_{22}^*\rangle =(1/2)-(1/2)\exp(-z/L_{fwm})$, $\langle u_{10}u_{18}\rangle =-\exp(-z/L_{fwm})$, and
$\langle u_{10}u_{18}^*\rangle =(1/2)+(1/2)\exp(-z/L_{fwm})$.

Finally, matrix $M_C$ possesses a nondegenerate eigenvalue $\tilde\lambda_C=0$ in the limit $\Delta^{(-)}=0$, with the other eigenvalues vanishing for 
$z\ge L_{transient}$. The perturbative approach yields the correction to the zero eigenvalue: $\lambda_C=-L_d^{-1}$. Then, for $L\ge L_{transient}$ we 
find $\langle u_6u_{29}\rangle =0$, $\langle u_6u_{29}^*\rangle =(2/3)\exp(-z/L_d)$, 
$\langle u_3u_{25}\rangle =\langle u_3u_{25}^*\rangle =(1/3)\exp(-z/L_d)$.

When all these nonlinear coefficients are substituted in the equations for the field, we arrive to the final result: Eqs.~(\ref{3})-(\ref{5}), which represent
equations with deterministic coefficients, as desired.

\newpage
\section*{Figure Captions}
\begin{figure}[h]
\includegraphics[width=0.55\textwidth]{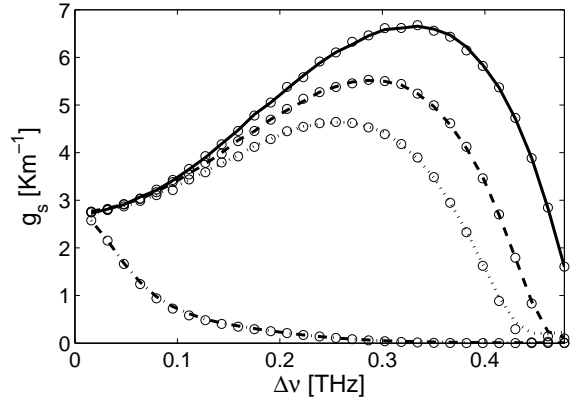}
\caption{Dependence of signal gain $g_s$ on its frequency detuning from the pump, with $L=300$~m. Curves and circles were obtained with 
$z$-varying or average $M$ coefficients, respectively. Moreover $\Delta n=0$ (solid curve); $\Delta n=1.0\cdot10^{-6}$ (dashed curve);
$\Delta n=1.5\cdot10^{-6}$ (dotted curve); $\Delta n=1.0\cdot10^{-5}$ (dash-dotted curve). }
\label{f1}
\end{figure}

\begin{figure}[h]
\includegraphics[width=0.55\textwidth]{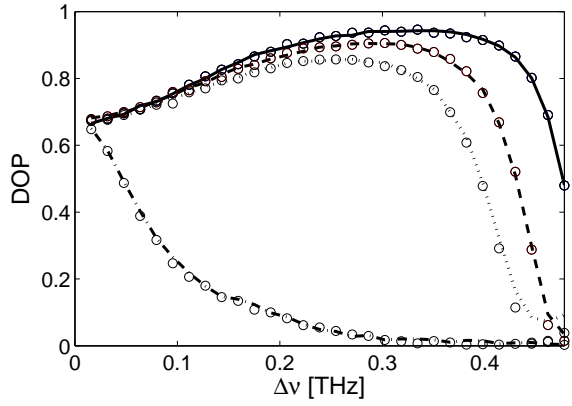}
\caption{Same as Fig.\ref{f1}, but for the signal DOP}
\label{f2}
\end{figure}

\begin{figure}[h]
\includegraphics[width=0.55\textwidth]{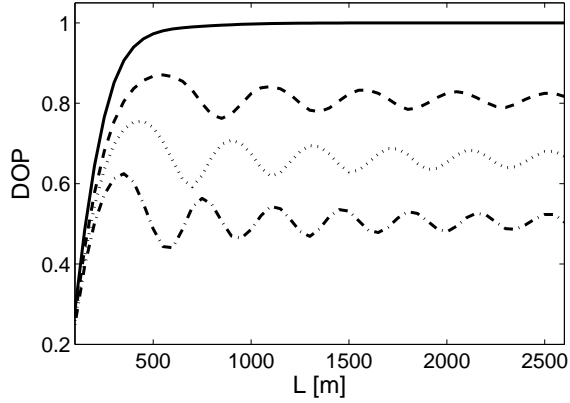}
\caption{Signal DOP versus fiber length $L$ with $\Delta n=1.5\cdot10^{-6}$, and different values of the sideband detuning frequency: 
$\Delta\nu =0.255$~THz (solid curve); $\Delta\nu =0.350$~THz (dashed curve); $\Delta\nu =0.365$~THz (dotted curve); $\Delta\nu =0.380$~THz 
(dash-dotted curve).}
\label{f3}
\end{figure}

\begin{figure}[h]
\includegraphics[width=0.45\textwidth]{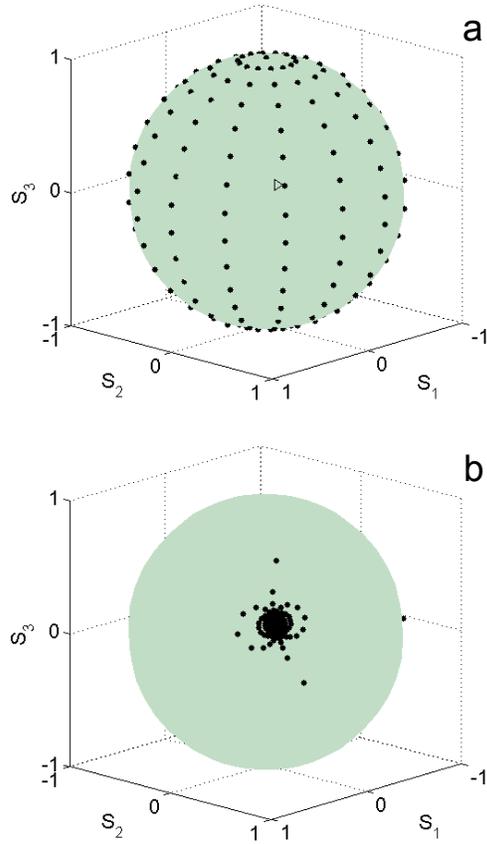}
\caption{Tips of input (a) and output (b) signal Stokes vectors on the Poincar\'e sphere for a fiber length $L=500$~m, $\Delta n=1.5\cdot10^{-6}$, and
$\Delta_\nu =0.255$~THz. For the sake of clarity, only $225$ vectors are represented instead of the $10000$ used in the simulations. Input vectors are
distributed uniformly over the sphere. The empty triangle represents the input pump Stokes vector.}
\label{f4}
\end{figure}

\begin{figure}[h]
\includegraphics[width=0.45\textwidth]{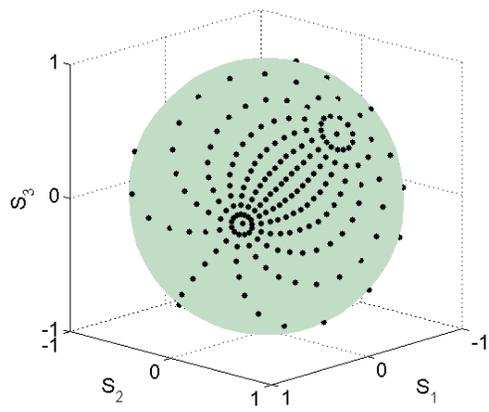}
\caption{Output signal Stokes vectors with $L=500$~m and $\Delta\nu =0.350$~THz.}
\label{f5}
\end{figure}

\begin{itemize}

\item Figure 1: Dependence of signal gain $g_s$ on its frequency detuning from the pump, with $L=300$~m. Curves and circles were obtained with 
$z$-varying or average $M$ coefficients, respectively. Moreover $\Delta n=0$ (solid curve); $\Delta n=1.0\cdot10^{-6}$ (dashed curve);
$\Delta n=1.5\cdot10^{-6}$ (dotted curve); $\Delta n=1.0\cdot10^{-5}$ (dash-dotted curve).

\item Figure 2: Same as Fig.\ref{f1}, but for the signal DOP.

\item Figure 3: Signal DOP versus fiber length $L$ with $\Delta n=1.5\cdot10^{-6}$, and different values of the sideband detuning frequency: 
$\Delta\nu =0.255$~THz (solid curve); $\Delta\nu =0.350$~THz (dashed curve); $\Delta\nu =0.365$~THz (dotted curve); $\Delta\nu =0.380$~THz 
(dash-dotted curve).

\item Figure 4: Tips of input (a) and output (b) signal Stokes vectors on the Poincar\'e sphere for a fiber length $L=500$~m, $\Delta n=1.5\cdot10^{-6}$,
and $\Delta_\nu=0.255$~THz. For the sake of clarity, only $225$ vectors are represented instead of the $10000$ used in the simulations. Input vectors 
are distributed uniformly over the sphere. The empty triangle represents the input pump Stokes vector.

\item Figure 5: Output signal Stokes vectors with $L=500$~m and $\Delta\nu =0.350$~THz.

\end{itemize}

\newpage

\end{document}